\documentclass{article}
\usepackage{arxiv}

\usepackage[utf8]{inputenc} % allow utf-8 input
\usepackage[T1]{fontenc}    % use 8-bit T1 fonts
\usepackage{hyperref}       % hyperlinks
\usepackage{url}            % simple URL typesetting
\usepackage{booktabs}       % professional-quality tables
\usepackage{amsfonts}       % blackboard math symbols
\usepackage{amsmath}        % blackboard math symbols
\usepackage{amssymb}        % blackboard math symbols
\usepackage{nicefrac}       % compact symbols for 1/2, etc.
\usepackage{microtype}      % microtypography
\usepackage{lipsum}
\usepackage{enumitem}
\usepackage{multicol}
\usepackage{graphicx}
\usepackage{float}
\usepackage{comment}
\usepackage[all]{hypcap}
\hypersetup{
colorlinks=true,
linkcolor=blue,
citecolor=blue,
urlcolor=blue
}

\usepackage[
        retain-explicit-plus=true,
        retain-unity-mantissa=false,
        range-phrase=--,
        range-units=single
    ]{siunitx}

\newcommand{\beq}[1]{\begin{equation}\label{eqn:#1}}
\newcommand{\eeq}[0]{\end{equation}}

\def\SCoV{SARS-CoV-2}
\def\COV{COVID-19}

\def\ssim{\sim\!}
\newcommand{\VarWithSub}[2]{{\ensuremath{#1_{\mathrm{#2}}}}}
\newcommand{\VarWithSubSup}[3]{{\ensuremath{#1_{\mathrm{#2}}^{\mathrm{#3}}}}}

\def\Pinf{\VarWithSub{P}{inf}}
\def\Dinf{\VarWithSub{D}{inf}}
\def\DinfSS{\VarWithSubSup{D}{inf}{SS}}
\def\DinfSys{\VarWithSubSup{D}{inf}{Sys}}
\def\DinfTE{\VarWithSubSup{D}{inf}{TE}}
\def\DinfCE{\VarWithSubSup{D}{inf}{CE}}
\def\DinfPS{\VarWithSubSup{D}{inf}{PS}}
\def\DinfNom{\VarWithSubSup{D}{inf}{0}}

\def\tmaxSS{\VarWithSubSup{t}{max}{Room}}
\def\tmaxSys{\VarWithSubSup{t}{max}{Sys}}
\def\tminTE{\VarWithSubSup{t}{min}{TE}}
\def\tminPS{\VarWithSubSup{t}{min}{PS}}

\def\Ninf{\VarWithSub{N}{inf}}

\def\rsrc{\VarWithSub{r}{src}}
\def\Vsrc{\VarWithSub{V}{src}}
\def\rroom{\VarWithSub{r}{room}}
\def\Vroom{\VarWithSub{V}{room}}
\def\rsys{\VarWithSub{r}{sys}}
\def\Vsys{\VarWithSub{V}{sys}}
\def\fHVAC{\VarWithSub{f}{sys}}

\def\rhoA{\VarWithSub{\rho}{A}}
\def\NA{\VarWithSub{N}{A}}
\def\rhoASS{\VarWithSubSup{\rho}{A}{SS}}

\def\tauroom{\VarWithSub{\tau}{room}}
\def\tausys{\VarWithSub{\tau}{sys}}
\def\taudecay{\VarWithSub{\tau}{decay}}
\def\taufall{\VarWithSub{\tau}{fall}}

\def\usettle{\VarWithSub{u}{settle}}
\def\Aroom{\VarWithSub{A}{room}}

\def\ACH{\ensuremath{\rm ACH}}

\title{Avoiding COVID-19: Aerosol Guidelines}

\author{
  Matthew J.~Evans \\ % \thanks{Use footnote for providing further information about author} \\
  Department of Physics\\
  Massachusetts Institute of Technology\\
  Cambridge, MA 02139 \\
  \texttt{m3v4n5@mit.edu} \\
}

\begin{document}
\maketitle

\begin{abstract}
The \COV\ pandemic has brought into sharp focus the need to understand
 respiratory virus transmission mechanisms.
In preparation for an anticipated influenza
 pandemic, a substantial body of literature has developed over the last
 few decades showing that the short-range aerosol route
 is an important, though often neglected transmission path.
We develop a simple mathematical model for \COV\ transmission
 via aerosols, apply it to known outbreaks,
 and present quantitative guidelines
 for ventilation and occupancy in the workplace.
\end{abstract}

% keywords can be removed

%\vskip 0.2in

\begin{multicols}{2}

%%%%%%%%%%%%%%%%%%%%%%%
\section{Introduction}

The world is learning to navigate the \COV\ pandemic and
 a great deal of information about the disease is already available
\cite{Harvard_COVID_course,CDC_COVID,10.1001/jama.2020.3072}.
In order to adjust to this new reality
 it is important to understand what
 can be done to avoid infection, and to avoid infecting others.
 
\SCoV, the coronavirus which causes \COV, is thought to be transmitted
 via droplets, surface contamination and aerosols
\cite{Asadi:2020bv, Wang:2020we, Anderson:2020kx, Dietz:2020im, Meselson:2020xx, CDC_spread,ASHRAE_aerosols}.
\COV\ outbreaks are most common in indoor spaces
 \cite{Qian2020.04.04.20053058}, 
 and evidence points to a dominant role of aerosol transmission
 \cite{Ferretti:2020wd, Morawska:2020wa}.
Furthermore, infection via inhalation of aerosols
 and small droplets dominates
 large droplet exposure in most cases for physical reasons linked to
 complex, but well understood fluid dynamics
 \cite{Bourouiba2014, ShortRangeChen2020, ShortRangeLiu2020, BourouibaRev2020}.
Despite this fact, there is a common misconception that aerosol
 transmission implies efficient long-range transmission (as in measles),
 and thus that the absence of long-range transmission implies
 an absence of aerosol transmission.
The truth is more nuanced, and includes the possibility of a dominant
 short-range aerosol path limited by pathogen dilution, deposition,
 and decay \cite{Buonanno:2020vc,Tellier:2006ww,ASHRAE_airborne}.

The production of aerosolized viruses by a contagious individual occurs
 naturally as a result of discrete events (e.g., coughing or sneezing),
 or through a continuous process like breathing and talking
  \cite{Buonanno:2020wj, Mittal:2020jz}.
Droplets that are formed with diameters less than
 $\ssim \SI{50}{\mu m}$ quickly
 lose most of their water to evaporation, shrinking by a factor
 of 2-3 in diameter and becoming ``droplet nuclei''.
These fine particles settle slowly
 and mix with the air in the environment
 \cite{EvapXie2007, Somsen:2020fa, Stadnytskyi:2020vn}.

Aerosol transmission happens when a susceptible individual
 inhales these now sub-\SI{20}{\mu m} droplet nuclei
 that are suspended in the air around them \cite{Tellier2009, KUTTER2018142}.
This is thought to be the dominant transmission
 mechanism for influenza and rhinovirus \cite{Tellier2009, 10.1093/infdis/156.3.442} and possibly also for SARS and MERS \cite{Tellier:2019vc}.
For influenza, it has been shown that aerosol particles as small as
 \SI{1.5}{\mu m} are sufficient for transmission \cite{ZhouE2386}.
Furthermore, it is known from other viral respiratory
 diseases that aerosol exposure can result in infection
 and illness at much lower doses than other means (e.g., nasal inoculation)
 \cite{TEUNIS2010215}.

In line with current recommendations \cite{CDC_prevention},
 we will assume that hand-hygiene protocols are being followed
 such that surface contamination
 remains a sub-dominant transmission route
 \cite{Ferretti:2020wd, CDC_spread}.
We will also assume that contagious individuals
 are wearing some kind of face covering which is sufficient to
 disrupt the momentum of any outgoing airflow
 \cite{Tang2009, viola2020face}
 and catch large droplets
 \cite{Rodriguez-Palacios2020.04.07.20045617}.
These actions ensure that the aerosols investigated here
 remain the dominant transmission mechanism.

In the following sections we present a simple model for
 aerosol transmission in a variety of contexts,
 apply this model to known outbreaks,
 and develop guidelines for reducing the probability of
 transmission in the workplace.

%%%%%%%%%%%%%%%%%%%%%%%
\section{Risk Assessment}
\label{sec:risk}

In this section we motivate the analysis that follows
 as a foundation for guidelines based on the
 the probability that a contagious individual
 will infect a coworker, $\Pinf$,
 and the absolute risk of infection during an epidemic
 such as \COV.
The later of these is not addressed directly by the analysis
 presented in this work and depends on the prevalence of
 highly-contagious (and presumably asymptomatic or pre-symptomatic)
 individuals in the population \cite{He2020.03.15.20036707,Wolfel:2020ti,doi:10.1056/NEJMc2001468, Aguilar2020.03.18.20037994, Cheng2020.03.18.20034561}.
In the context of the current \COV\ pandemic, a return to
 work seems improbable if more than 1 in 1000 people are
 unwittingly contagious on any given day.  This suggests that
 the \emph{absolute probability} of infection should
 be less than $\Pinf / 1000$, since our calculations of
 $\Pinf$ \emph{assumes the presence of a contagious individual}.
An acceptable workplace transmission risk should be chosen such
 that a contagious employee has only a modest chance
 of infecting another employee during their pre-symptomatic
 period, and such that the absolute risk of infection
 remains acceptably low (e.g., less than 0.01\%
 if $\Pinf \lesssim 10\%$).

A more quantifiable and meaningful approach to risk assessment can be
 derived from the relative risk of infection by \COV\ in the
 workplace vs. other locations, and the associated contribution
 of the workplace to the growth or decline of the epidemic
 (i.e., the reproduction number $R_0$) \cite{Howard:2020it}.
The grander objective of ensuring that the 
 workplace's contribution to the epidemic's $R_0$ is small
 is sufficient to ensure that the relative risk of contracting \COV\ in
 the workplace is small compared with other activities
 (e.g., homelife or recreation).
A workplace transmission probability of 10\%, for instance,
 would contribute 0.1 to $R_0$,
 meaning that for a decaying epidemic with $0.5 < R_0 < 1$,
 an employee would be 5 to 10 times more likely to be infected
 away from work than at work.

This relative view of risk assessment requires a somewhat
 counter-intuitive approach to setting guidelines:
 the focus should not be on limiting the probability that
 an employee becomes infected, but rather on limiting the
 quantity of pathogens \emph{delivered to others}
 by a contagious and as-yet-asymptomatic employee.
 
%%%%%%%%%%%%%%%%%%%%%%%
\section{Infection Model}
\label{sec:infection}

The probability of developing an infection \Pinf\
 given exposure to a volume $V$ of saliva with a concentration $\rho$
 of virus particles (``virions'') is 
\beq{Pinf}
\Pinf = 1 - e^{-\rho \, V / \Ninf}
\eeq
 where the infectivity $\Ninf$ is the number of virions needed to make infection likely
 in the lungs for aerosol inhalation
 \cite{Buonanno:2020wj, TEUNIS2010215, Watanabe:2010el,Noakes:2009km,ASHRAE_airborne}.
The infectivity value $\Ninf$ includes probable deposition location
 (i.e., upper vs. lower respiratory tract) and local deposition
 efficiency (e.g., not all inhaled droplets
 are deposited in the lungs
 \cite{TEUNIS2010215, LungDep_Hatch1961,ASHRAE_LungDep}).
It should be noted, however, that small doses are
 less likely to cause \emph{illness} than what is indicated by
 Eqn. \ref{eqn:Pinf} \cite{TEUNIS2010215, Nicas2006}.

Rather than work explicitly with the probability of infection \Pinf,
 we will use the ratio of the viral dose and the infectivity
\beq{Dinf}
\Dinf = \rho \, V / \Ninf
\eeq
 as a proxy, and refer to it as the ``infectious dose''.
Note that the infectivity and the viral concentration always
 appear together here, so only their ratio $\rho / \Ninf$
 is important in our model.
While $\rho$ has been measured for \COV\ \cite{Wolfel:2020ti, Jones2020},
 $\Ninf$ has not been directly measured.
We estimate $\Ninf \ssim 1000$ from influenza and other coronaviruses \cite{Watanabe:2010el, Nikitin:2014ua},
 and check this value by applying our model to known outbreaks
 (see appendix \ref{sec:outbreaks}).

The relationship between $\Dinf$ and the probability of
 infection $\Pinf$ is shown in Fig. \ref{fig:Dinf} (right).
Equation \ref{eqn:Pinf} is shown as the dashed-grey curve,
 indicating that the probability of infection is high
 for any infectious dose greater than 1,
 and $\Pinf$ is essentially proportional to $\Dinf$
 when $\Dinf \ll 1$.
The other curves in Fig. \ref{fig:Dinf} account for
 variability in $\rho$, $V$ and $\Ninf$, as described
 in section \ref{sec:ptrn}.

%%%%%%%%%%%%%%%%%%%%%%%
\section{Aerosol infectious Dose Model}
\label{sec:dose}

In this section we compute the infectious dose which results from
 the presence of a contagious individual in various scenarios.
This informs prescriptions for the maximum time
 a susceptible individual may be exposed to potentially
 contaminated air, or the minimum time between occupants in
 the same space.
Each of these calculations uses the values in Table \ref{tab:parameters}
 and concludes with an exposure time limit for
 a given infectious dose.
The infectious dose limits used here
 are chosen to maintain a transmission probability
 less than 10\%, as motivated in section \ref{sec:risk}
 with examples in section \ref{sec:ptrn}.

%%%%%%%%%%%%%%%%%%%%%%%
\begin{table*}
\label{tab:parameters}
  \centering
  \begin{tabular}{llccc}
    \toprule
    Description     & Symbol     & Value &  Reference \\

    \midrule %%%%%%%%%%%%%%%%%%%%%%
    Viral Load in Saliva & $\rho_0$  & \SI{1000}{/ n L}  &
     \cite{Wolfel:2020ti, Jones2020, To:2020vv} \\
    Sneeze Aerosol Volume & $V_s$  & \SI{1}{\mu L}  &
     \cite{duguid_1946, WHO_droplets} \\
    Cough Aerosol Volume & $V_c$  & \SI{100}{n L}  &
     \cite{duguid_1946, WHO_droplets} \\
    Talking Aerosol Volume & $V_t$  & \SI{10}{n L / min}  &
     \cite{duguid_1946, WHO_droplets} \\
    Breathing Aerosol Volume & $V_b$  & \SI{1}{n L / min}  &
     \cite{duguid_1946, WHO_droplets} \\
    Aerosol Decay Time & $\tau_a$ & $\sim \SI{20}{min}$ &
     see Sec. \ref{sec:taua} \\
     %\cite{Somsen:2020fa, Stadnytskyi:2020vn, VanDoremalen2020, Fears2020} \\
    Breathing Rate & $r_b$ & \SI{10}{L/min} &
     \cite{Tidal2020} \\      
    Respiratory Infectivity & $\Ninf$ & 1000 & 
     \cite{Watanabe:2010el, TEUNIS2010215} \\ 
    \midrule %%%%%%%%%%%%%%%%%%%%%%
    \multicolumn{2}{c}{Frequently Used Symbols} &  Units & Equation \\
    \midrule %%%%%%%%%%%%%%%%%%%%%%
    Infectious Dose (general) & $\Dinf$ & number & Eqn. \ref{eqn:Dinf} \\
    Time & $t$ & \si{min} & Eqn. \ref{eqn:drhodt} \\
    Aerosol Source Rate & $\rsrc$  & \si{nL / min} &
     Eqn. \ref{eqn:drhodt} \\
    Aerosol Source Volume & $\Vsrc$ & \si{nL} &
     Eqn. \ref{eqn:DinfTE} \\
    Room Volume & $\Vroom$ & \si{m^3} &
     Eqn. \ref{eqn:tauroom} \\
    Room Ventilation Rate & $\rroom$ & \si{m^3 / min} &
     Eqn. \ref{eqn:tauroom} \\
    Room Air-Cycle Time & $\tauroom$ & \si{min} &
     Eqn. \ref{eqn:tauroom} \\
    Aerosol Decay Factor & $f_a$ & number &
     Eqn. \ref{eqn:fa} \\
    \midrule %%%%%%%%%%%%%%%%%%%%%%
    \multicolumn{4}{c}{Conversion Factors}  \\
    \midrule %%%%%%%%%%%%%%%%%%%%%%
   Cubic Feet per Minute & CFM &
    $\SI{1}{m^3 / min} \simeq \SI{35}{CFM}$  \\
   Air Changes per Hour & ACH & $\tauroom \simeq \SI{60}{min} / \ACH$  \\
   \midrule %%%%%%%%%%%%%%%%%%%%%%
    &\multicolumn{3}{c}{Maximum Exposure and Minimum Wait Times}                   \\
    \cmidrule(r){2-4}
    %%%
    Room Steady State & \multicolumn{2}{l}
    {$\tmaxSS \simeq \SI{100}{min} \; \frac{\SI{1}{n L / min}}{\rsrc} \; 
 \frac{\rroom}{\SI{10}{m^3 / min}} $} & Eqn. \ref{eqn:tmaxSS} \\
    %%%
  System Steady State & \multicolumn{2}{l}{$\tmaxSys \simeq \SI{5}{hour} \;
    \frac{0.1}{1 - \fHVAC} \; \frac{\SI{3}{n L / min}}{\rsrc} \; 
 \frac{\fHVAC \rsys}{\SI{100}{m^3 / min}}$} & Eqn. \ref{eqn:tmaxSys} \\
  %%%
    Occupancy Wait Time & \multicolumn{2}{l}{$\tminTE \simeq \tauroom \;
 \ln \left(10 \; \frac{\Vsrc}{\SI{100}{n L}} \; 
 \frac{\SI{1}{m^3 / min}}{\rroom} \right)$} & Eqn. \ref{eqn:tminTE} \\
  %%%
    Passage Wait Time & \multicolumn{2}{l}{$\tminPS \simeq \tauroom \;
 \ln \left(10 \; \frac{\Vsrc}{\SI{100}{n L}} \; 
 \frac{\SI{10}{m^3}}{\Vroom} \frac{\Delta t}{\SI{1}{min}}\right)$} & Eqn. \ref{eqn:tminPS} \\
    \bottomrule
  \end{tabular}
  \caption{Parameters used in the aerosol transmission model.
    See section \ref{sec:parameters} for more information.}
%\vspace{2ex}
\end{table*}

%%%%%%%%%%%%%%%%%%%%%%%
\subsection{Steady State in a Room}
\label{sec:steady}

A contagious individual will shed virions into the room they occupy 
 through breathing, talking, coughing, etc.
The aerosolized virion concentration in a room can be expressed in the form of
 a differential equation as
\beq{drhodt}
\frac{d \rhoA(t)}{dt} = \rho_0 \frac{\rsrc}{\Vroom} -
  \rhoA(t) \left( \frac{1}{\tauroom} + \frac{1}{\tau_a} \right)
\eeq
 where $\rho_0$ is the nominal viral concentration in saliva,
 which is emitted in aerosol form at a rate of $\rsrc$,
 and $\tau_a$ is the timescale for aerosol concentration decay.
The air-cycle time in a room is
\beq{tauroom}
 \tauroom = \Vroom / \rroom
\eeq
 where $\rroom$ is rate at which air is removed from the room (or filtered locally).
The steady-state concentration is
\beq{rhoSS}
\rhoASS = \rho_0 \; \frac{\rsrc}{\Vroom} \;
\frac{\tauroom \tau_a}{\tauroom + \tau_a}
\eeq
 and has units of viral copies per liter of air.

Some simplification can be achieved by using Eqn. \ref{eqn:tauroom}
 to rewrite Eqn. \ref{eqn:rhoSS} as
\beq{rhoSS2}
\rhoASS = \rho_0 \; \frac{\rsrc}{\rroom} \; f_a
\eeq
 where
\beq{fa}
f_a = \frac{\tau_a}{\tauroom + \tau_a} \sim \frac{\ACH}{\ACH + 3}
\eeq
 is the aerosol concentration decay factor
 which results from virion decay and setting of droplet nuclei
 (see section \ref{sec:taua}).
The approximation gives $f_a$ in terms of air-changes
 per hour (ACH) for a 3 meter ceiling height.
In the well ventilated limit (i.e., $\tauroom \ll \tau_a$)
 $f_a$ goes to 1, meaning that settling is not relevant.
In poorly ventilated spaces (i.e., $\tauroom \gtrsim \tau_a$),
 however, settling can play an important role
 since it effectively adds 3 air-changes per hour
 to the rate of aerosol concentration decay.

Breathing air in a room with an aerosol virion concentration
 $\rhoA$ will cause an
 accumulation of exposure $\NA$ (number of virions)
 proportional to the time passed in that room $t$
\beq{NA}
\NA = \rhoA \; r_b t
\eeq
 such that the steady-state infectious dose is
\beq{DinfSS}
\DinfSS =  \frac{\rho_0}{\Ninf} \; \frac{\rsrc}{\rroom} \;
 f_a \; r_b t
\eeq
 where $r_b \ssim \SI{10}{L/min}$ is the breathing rate. 
If the room is well ventilated (i.e., $f_a \ssim 1$)
 the infectious dose is
\beq{DinfSS2}
 \DinfSS \simeq 0.01 \; \frac{\rsrc}{\SI{1}{n L / min}} \; 
 \frac{\SI{1}{m^3 / min}}{\rroom} \; \frac{t}{\SI{1}{min}}
\eeq
 for any space (e.g. office, laboratory or bathroom)
 occupied by a contagious individual.

The infectious dose crosses our example event exposure
 threshold of 0.1 in an office-like space
 ($\rroom \ssim \SI{10}{m^3 / min}$)
 after 100 minutes,
\beq{tmaxSS}
 \tmaxSS \simeq \SI{100}{min} \; \frac{\SI{1}{n L / min}}{\rsrc} \; 
 \frac{\rroom}{\SI{10}{m^3 / min}} \;,
\eeq
 while in small spaces with less ventilation
 (e.g., a car or bathroom with $\rroom \ssim \SI{1}{m^3 / min}$)
 the dose would cross the threshold after only \SI{10}{minutes}.
This assumes that the contagious individual is quietly working
 such that $\rsrc = V_b$,
 but if they are talking or coughing occasionally
 the source term may be much higher (e.g., $\rsrc \!\gg\! V_b$).
The implication of this example is that in order to share a space
 for 8 hours while maintaining $\DinfNom < 0.1$, 
 $\rroom$ should be greater than
  $\SI{50}{m^3 / min} \ssim \SI{2000}{CFM}$
  per occupant beyond the first,
 making shared occupancy of typical office spaces untenable.
Allowing for a higher nominal dose (e.g., $\DinfNom \ssim 1$),
 which may be acceptable
 if the incidence of \COV\ in the population is low,
 would reduce this requirement by a factor of 10.
 
%%%%%%%%%%%%%%%%%%%%%%%
\subsection{Steady State in a Building}
\label{sec:HVAC}

When a susceptible individual occupies a room that is
 connected via the HVAC system to a room occupied by
 a contagious individual, there is the possibility of
 aerosol transmission through the HVAC system \cite{Dietz:2020im, Li:2007gj,CDC_air}.
 
However, the volume of a droplet is proportional to
 its diameter \emph{cubed}, which leads to the majority
 of the virions being delivered in the $\ssim \SI{10}{\um}$ droplet nuclei
 \cite{EvapXie2007, ZhouE2386}, which are filtered by HVAC systems.
Most HVAC filters will remove the majority of the viral
 load associated with \COV\ transmission, and a high quality
 filter (i.e., MERV 12) will remove at least
 90\% of relevant droplet nuclei \cite{Elliott:2003vb}.
The remaining aerosols are further diluted by fresh make-up air,
 and then spread among all of the air spaces associated with
 the HVAC system.
The associated infectious dose in
 rooms connected to a room occupied by a contagious individual
 will be roughly
\beq{DinfSys}
\DinfSys = (1 - \fHVAC) \frac{\rho_0}{\Ninf} \; \frac{\rsrc}{\Vsys} \;
 \frac{\tausys \tau_a}{\tausys + \fHVAC \tau_a} \; r_b t
\eeq
 where $\fHVAC$ is the fraction
 of the virions which is removed by the filters
 or displaced by make-up air,
 and $\tausys = \rsys / \Vsys$ is the time required to cycle the entire
 air volume through the HVAC system.

In the well ventilated limit, as above, the infectious dose is
\beq{DinfSys2}
 \DinfSys \simeq 10^{-5} \; \frac{1 - \fHVAC}{0.1} \; 
 \frac{\SI{100}{m^3 / min}}{\fHVAC \rsys} \; \frac{\rsrc \, t}{\SI{1}{n L}}
\eeq
 and the maximum occupancy time for $\DinfSys < 0.01$ is
\beq{tmaxSys}
 \tmaxSys \simeq \SI{5}{hour} \; \frac{0.1}{1 \!-\! \fHVAC} \;
 \frac{\SI{3}{n L / min}}{\rsrc} \; 
 \frac{\fHVAC \rsys}{\SI{100}{m^3 / min}} \; .
\eeq
We have set the $\rsrc$ scale at \SI{3}{n L / min}
 to allow for occasional conversation (e.g., video conferencing).
The maximum dose used in this example is 0.01 instead of 0.1
 to allow for 10 exposed individuals
 (see section \ref{sec:ptrn}).

To understand why transmission through HVAC systems appears to 
 be rare we note that even medium quality air filters (MERV 9)
 provide better than 99\% filtering after some loading \cite{Elliott:2003vb}.
The value we use here assumes little or no fresh make-up air
 and is representative of an unloaded (i.e., clean)
 air filter, which is the \emph{worst case}.
Also, in buildings where the HVAC system does not
 recirculate air this type of transmission cannot occur
 (i.e., $f_{sys} = 1$).
 
\begin{figure*}[ht!]
    \label{fig:Dinf}
 \begin{center}
 \includegraphics[width=0.48\textwidth, trim=15mm 5mm 15mm 18mm, clip]{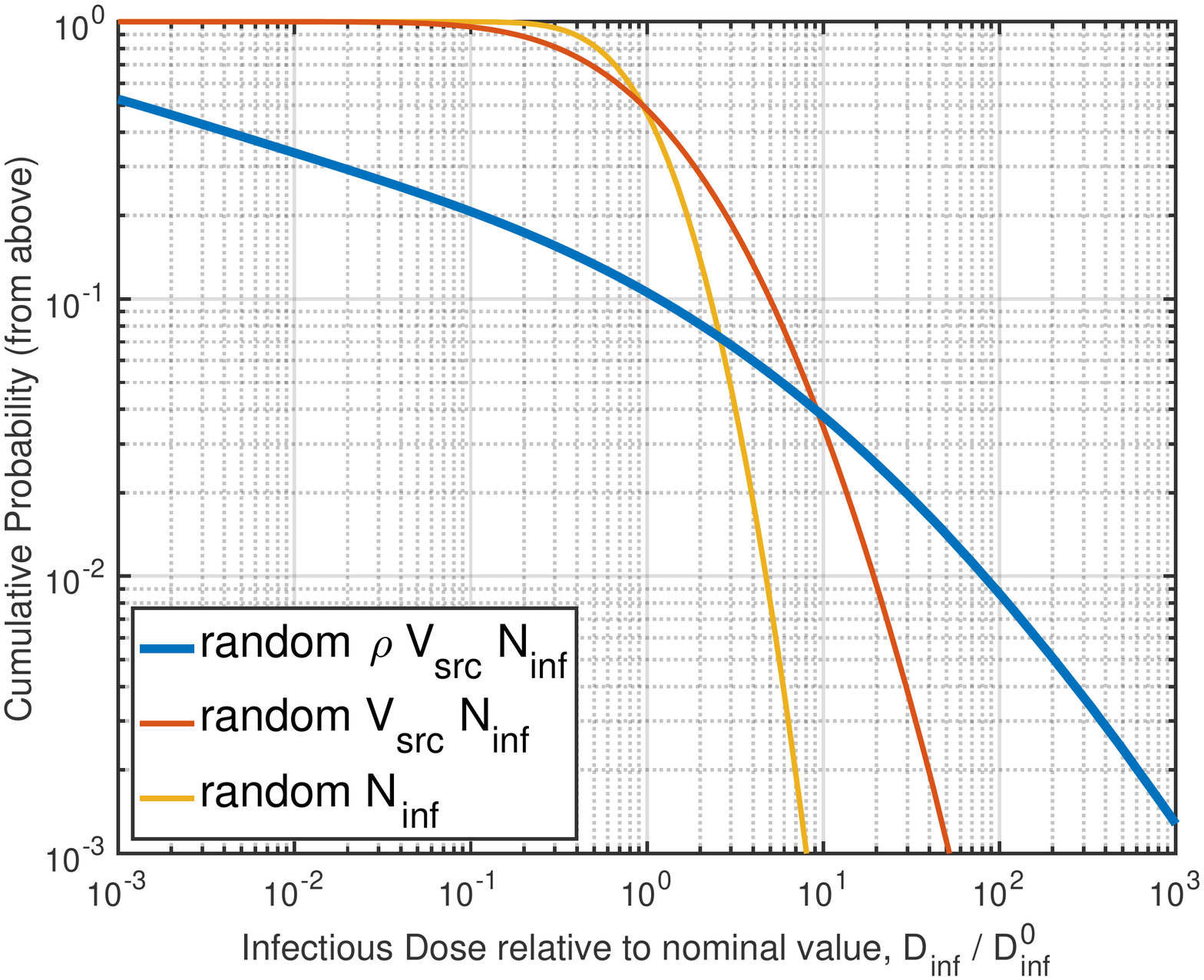}  % [trim=left bottom right top, clip]
 \hspace{10pt}
 \includegraphics[width=0.48\textwidth, trim=15mm 5mm 15mm 18mm, clip]{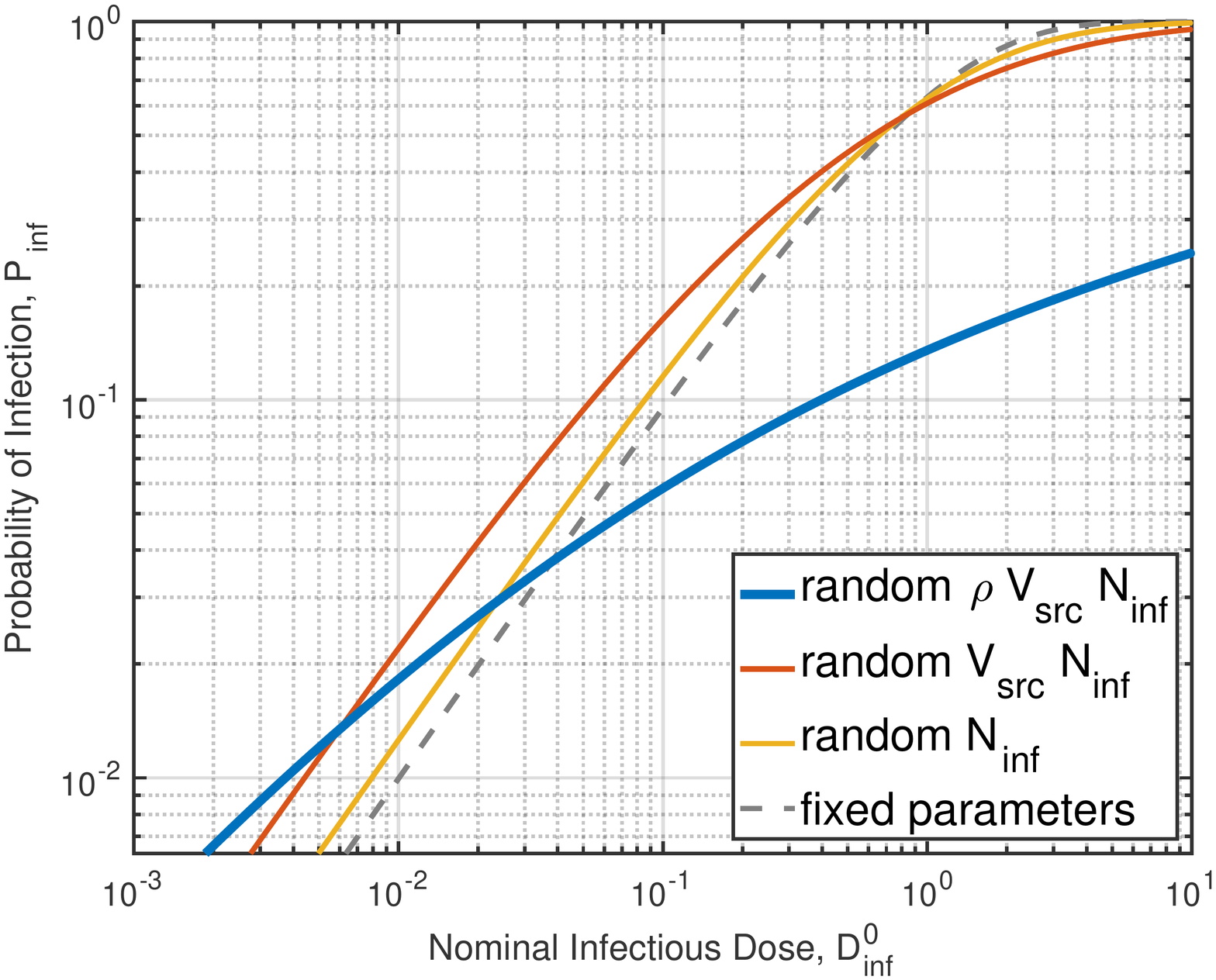}
 \end{center}
 \caption{\textbf{Left}: Probability that the infectious dose in any
  given encounter exceeds the nominal dose by some factor.
 The three curves allow some or all transmission parameters to vary randomly
  (see section \ref{sec:parameters} and appendix \ref{sec:pdfs}).
 For instance, allowing all parameters to vary (blue trace)
  the probability of having $\Dinf > \DinfNom$ is about 10\%,
  $\Dinf > 10 \, \DinfNom$ will happen in about 4\% of cases,
  while 1\% of cases will have $\Dinf \gtrsim 100 \, \DinfNom$.
Choosing the nominal viral load $\rho_0$ equal to the median value
 of \SI{1}{/ n L} (rather than the $90^{th}$ percentile as we have)
 would shift the blue curve to the right by a factor of 1000.
  \hspace{1ex}
  \textbf{Right}: Probability of infection as a function of 
   nominal infectious dose $\DinfNom$.
  The three solid curves relate $\Pinf$ to $\DinfNom$ allowing some
   or all transmission parameters to vary randomly,
   while the dashed curve (grey) shows $\Pinf$ with fixed parameters
   as in Eqn. \ref{eqn:Pinf}.
Allowing all parameters to vary, the probability of infection for
  a nominal dose of 1 is 14\%.  $\DinfNom = 0.1$
  reduces $\Pinf$ to 6\%, and
  $\DinfNom \lesssim 3 \times 10^{-3}$ is required for
  $\Pinf < 1\%$.
Choosing the nominal viral load $\rho_0$ equal 
 to \SI{1}{/ n L} rather than \SI{1000}{/ n L} would reduce
 $\DinfNom$ by 1000
 relative to our calculations in section \ref{sec:dose} and shift
 the blue curve to the left by 1000,
 such that $\Pinf$ would remain unchanged.
The ``random $\Vsrc \Ninf$''
  curve is used in appendix \ref{sec:outbreaks} to evaluate
  $\Pinf$ in cases where $\rho$ has a fixed value.}
\vspace{2ex}
\end{figure*}

As an explicit example, we consider 10 offices that
 are connected to a HVAC system
 which moves $\rsys = \SI{100}{m^3/min}$ with 30\% make-up air
 and a filter that removes 90\% of the virions
\beq{fsys}
\fHVAC = 1 - (1 - 0.3) (1 - 0.90) = 0.93 \; .
\eeq
We assume that one of these offices is occupied by
 a contagious individual who is emitting
 $\rsrc = \SI{3}{n L / min}$ of saliva in
 small droplets which evaporate to form airborne droplet nuclei.
According to Eqn. \ref{eqn:DinfSys2},
 each of the individuals in the other offices will receive an
 infectious dose of 0.01 after working for 8 hours.
The total dose, summed over all susceptible individuals
 and assuming that all offices are occupied, is 0.1,
 indicating that there is a 6\% chance that at least
 one of these individuals will be infected
 (see Fig. \ref{fig:Dinf}, right).
Generalizing this example, we see that
 $\SI{10}{m^3 / min} \ssim \SI{350}{CFM}$
 per occupant, averaged over all rooms connected by
 a well filtered HVAC system, is sufficient to prevent
 a large nominal dose.
% It is interesting to note that this probability does
%  not change with more \emph{occupied} offices
%  (e.g., 30 offices and $\rsys = \SI{300}{m^3/min}$)
%  since this will decrease the individual
%  dose while keeping the total dose unchanged.
% If, on the other hand, some offices are left \emph{unoccupied}
%  the probability of transmission decreases.

%%%%%%%%%%%%%%%%%%%%%%%

%%%%%%%%%%%%%%%%%%%%%%%
\subsection{Transient Occupancy and Events}
\label{sec:transient}

Some spaces are occupied briefly by many people (e.g., bathrooms)
 and may not have time to come to steady state.
Coughing and sneezing events can also cause a transient
 increase in the viral concentration in a room.
The infectious dose associated with transients
 can be quantified as
\beq{DinfTE}
\DinfTE =  \frac{\rho_0}{\Ninf} \; \frac{\Vsrc}{\Vroom} r_b \;
 \int_{t_1}^{t_2} e^{-t \frac{\tauroom + \tau_a}{\tauroom \tau_a}} dt
\eeq
 where the transient event occurs at time $t = 0$, 
 exposure is from time $t_1$ to $t_2$,
 and $\Vsrc$ is the aerosol source volume
 (e.g., $V_c$ for a cough, or $V_s$ for a sneeze).

We can estimate when a room is ``safe'' for a new occupant
 by computing the dose in the limit of $t_2 \rightarrow \infty$,
\beq{DinfTE2}
\DinfTE \simeq \frac{\rho_0}{\Ninf} \; \frac{\Vsrc}{\rroom}  \;
 f_a r_b  \; e^{-t_1 / f_a \tauroom} \; .
\eeq
This leads to a minimum wait time of
\beq{tminTE}
 \tminTE \simeq \tauroom \;
 \ln \left(10 \; \frac{\Vsrc}{\SI{100}{n L}} \; 
 \frac{\SI{1}{m^3 / min}}{\rroom} \right) \; ,
\eeq
 where we have set the $\Vsrc$ scale at \SI{100}{n L} to allow for
 some coughing from the previous occupant,
 and assumed good ventilation such that $f_a \ssim 1$.

As a concrete example, for a 70 square-foot
 bathroom with a 70 CFM fan in operation
 (i.e., $\tauroom \simeq \SI{10}{minutes}$ and $\rroom = \SI{2}{m^3 / min}$),
 Eqn. \ref{eqn:tminTE} gives a 16 minute wait time between occupants.
%Including the aerosol decay factor, $f_a = 2/3$,
% to get a more precise estimate reduces the wait time to 8 minutes.
Increasing the airflow to $\SI{10}{m^3 / min} \ssim \SI{350}{CFM}$ eliminates the wait time.

Going back to Eqn. \ref{eqn:DinfTE}, we can also compute the
 exposure due to brief passage through a common space
 in which $\Delta t = t_2 - t_1 \ll \tauroom$
 (e.g., corridors and elevators),
\beq{DinfPS}
\DinfPS \simeq \frac{\rho_0}{\Ninf} \; \frac{\Vsrc}{\Vroom}  \;
  r_b \Delta t \; e^{-t_1 / f_a \tauroom} \, .
\eeq
The minimum time between occupants for $\DinfPS < 0.01$ is
\beq{tminPS}
 \tminPS \simeq \tauroom f_a \;
 \ln \left(10 \; \frac{\Vsrc}{\SI{100}{n L}} \; 
 \frac{\SI{10}{m^3}}{\Vroom} \frac{\Delta t}{\SI{1}{min}}\right) \, ,
\eeq
 which will give negative values for large spaces or short
 passage times, indicating that no wait is necessary.
However, this assumes well mixed air, so some mixing time
 is necessary to avoid a ``close encounter''
 as described in the next section.
For small spaces with poor circulation
 long wait times are required:
 a sneeze in an elevator ($\Vsrc < \SI{1}{\mu L}$, 
 $\Vroom \ssim \SI{10}{m^3}$,
 $\Delta t \ssim \SI{1}{min}$)
 would put it out-of-order for 4.6 air changes.

A stairwell, for instance, with $\Vroom = \SI{100}{m^3}$
 and very little airflow ($\tauroom = \SI{60}{min}$
 and $f_a = 1/4$),
 would require a 34 minute wait between each use to allow
 airflow and droplet-nuclei settling to
 clear out the previous occupant's sneeze.
No waiting would be required if the previous occupant was
 just breathing or talking (even loudly) while in the stairwell
 (i.e., $\Vsrc < \SI{100}{n L}$).
However, stairwells require extra exertion during use,
 which will increase $r_b$, suggesting that at least
 10--15 minutes may be advisable before climbing many floors.
Note that while stairwells have unusually high ceilings,
 which could impact the time required for droplet nuclei to settle,
 the ratio of volume to horizontal surface area is
 similar to other spaces (see section \ref{sec:taua}
 for more discussion of settling times).

%%%%%%%%%%%%%%%%%%%%%%%
\subsection{Close Encounters}
\label{sec:close}

Mixing of aerosols into an airspace may require a few
 air-cycle times before a fairly uniform concentration
 can be assumed \cite{Hewlett:2013tt, anchordoqui2020physicist}.
In order to understand the potential infection risk associated
 with close proximity, we make a very rough estimate of
 the exposure in the vicinity of a mask-wearing cougher.
The cough momentum will be disrupted by the mask,
 but aerosols will exit the mask
 on essentially all sides \cite{viola2020face, Tang2009, Mittal:2020jz}
 resulting in a cloud around the cougher that will then
 either move with local air currents or rise with the
 cougher's body plume \cite{BodyPlumeCraven2006}.
We assume that large droplets are trapped in the
 cougher's mask or settle out around the cougher,
 while aerosols are dispersed into the air around the cougher.
The infectious dose at a distance $d$ for isotropic dispersion is
\beq{DinfCE}
\DinfCE = \frac{\rho_0 \, V_c}{\Ninf} \; \frac{3}{4 \pi \, d^3}  \; r_b t \; .
\eeq
For a typical cough this leads to
\beq{DinfCE2}
 \DinfCE \simeq 0.04 \; 
 \left( \frac{\SI{1}{m}}{d} \right)^3 \; 
 \frac{t}{\SI{10}{sec}}
\eeq
which suggests that an infection risk is still present at short distances,
 even if the cougher is wearing a mask \cite{10.7326/M20-1342}.
This analysis is clearly oversimplified, as details of the cough,
 the mask, and local air currents will
 all prevent isotropic dispersion,
 but serves to emphasize that extra care should be taken
 for transient events at short distances as they
 may result in a significant infectious dose.

%%%%%%%%%%%%%%%%%%%%%%%
\section{Mitigation Measures}
\label{sec:mitigation}

The calculations in the previous section allow a variety of
 possible mitigation measures.  This section briefly describes
 some means of reducing the probability of aerosol infection.
 
%%%%%%%%%%%%%%%%%%%%%%%
\subsection{Masks and Respirators}
\label{sec:masks}

Improvised face coverings, surgical masks and respirators,
 all collectively referred to as ``masks'' herein,
 serve a variety of purposes related to respiratory
 virus transmission:
\begin{enumerate}[leftmargin=*]
\item limiting direct transmission in face-to-face interactions
\item protecting the wearer by filtering inhaled air
\item protecting others by filtering exhaled air
\end{enumerate}

The first of these is the easiest to achieve:
 by disrupting any high-velocity outgoing
 pathogen-laden air flow \cite{viola2020face, Tang2009, Hui:2012dp, Bourouiba2014, BourouibaRev2020}
 and the capturing large droplets
 \cite{Rodriguez-Palacios2020.04.07.20045617, Davies:2013wo},
 masks of essentially any sort can limit direct
 transmission due to coughs, sneezes and conversations.
This paper assumes that this goal is achieved by
 universal mask use,
 thereby allowing for our simplified analysis of
 well-mixed aerosols.

Masks and respirators can reduce the probability of infection,
 both for the wearer and  for the people they interact with,
 by filtering the air inhaled and exhaled by the wearer
 \cite{Howard:2020it}.
While improvised face coverings
 generally do not remove more than half of 
 aerosolized particles on inhalation
 (protection factor PF $\lesssim 2$) \cite{Davies:2013wo},
 surgical masks provide some level of protection
 (PF of 2--10, typically around 3)
 \cite{MacIntyre2009, MacIntyreh694}.
Respirators which seal against the face are far
 more effective protection for the wearer
 (e.g., PF of 8--80 for N95s) \cite{10.1093/annhyg/men005},
 but benefiting from them requires proper fit and user compliance
 \cite{doi:10.1111/jebm.12381, doi:10.1080/15459620500330583,doi:10.1080/15459620490433799,peri2020analytical,CHEN202069,WANG2020},
 which suggests that widespread use is likely to be ineffective
 with currently available respirators
 \cite{MACINTYRE2008e328, Jefferson:2011vp, peri2020analytical}

Filtration of outgoing air is not well tested for most kinds
 of masks, but there is evidence of some protective
 value \cite{Davies:2013wo, viola2020face}.
Droplets produced while breathing, talking and coughing
 can be significantly reduced by mask usage
 \cite{Milton:2013ug,Leung:2020wf,Javidm1442,Mittal:2020jz},
 but much of the air expelled while coughing
 and sneezing goes around the mask \cite{viola2020face}.

Since effective filtration is difficult to achieve,
 both for incoming and outgoing air,
 we assign no protective value to the wearer for mask use,
 and assume no reduction in outgoing aerosols.
These are conservative assumptions which can be
 adjusted by reducing the infectious dose
 according to any assumed protection factor.
For example, a well fit N95 respirator should reduce
 the infectious dose received by the wearer by a factor of 20,
 and, in the absence of an exhale valve, may reduce the
 quantity of low-velocity outgoing aerosols by a similar factor.

%%%%%%%%%%%%%%%%%%%%%%%
\subsection{HVAC and Portable Air Filters}
\label{sec:filters}

Increasing air exchange rates in an HVAC system,
 and avoiding recirculation can both be used to
 reduce aerosol concentrations indoors.
Consistent use of local ventilation (e.g., bathroom fans)
 can also help to avoid infection.
In buildings and spaces where these measures are not
 available or not sufficient, local air filters
 (a.k.a., air scrubbers) can be used to increase
 the introduction of clean air into the space.
Small stand-alone units which filter \SI{10}{m^3 / min}
 or more are readily available and could help to increase
 safety in bathrooms and small offices.

As noted in sections \ref{sec:HVAC} and \ref{sec:parameters},
 the particle size of interest is greater than \SI{1}{\mu m},
 so special filtering technology is \emph{not} required \cite{Elliott:2003vb}.
Care should be taken, however, when changing filters
 both in stand-alone units and HVAC systems as they may
 contain significant viral load.
Stand-alone units should be disabled for at least 3 days
 before changing the filter to allow time for viral infectivity
 to decay \cite{VanDoremalen2020}.
Building HVAC filters should be changed with proper personal
 protective equipment, as recommended by the CDC \cite{Elliott:2003vb}.

%%%%%%%%%%%%%%%%%%%%%%%
\subsection{Clean Rooms and HEPA Filters}
\label{sec:HEPA}

Many laboratory spaces are outfitted with HEPA filters
 to provide clean air for laboratory operations.
Clean rooms offer a clear advantage over other spaces
 as they are designed to provide air which is free
 of small particles.

If the HEPA filters are part of a recirculating air
 cycle, Eqn. \ref{eqn:tmaxSys} can be used with
 $\fHVAC \gtrsim 0.999$, allowing for essentially
 indefinite exposure times.
It should be noted, however, that HEPA filters require
 lower air-speeds than those offered by typical HVAC
 systems and as such are not a ``drop in replacement''
 option.

Laboratories that use portable clean rooms in large
 spaces can be treated in a similar manner, using
 Eqn. \ref{eqn:DinfSys2} for the air inside the clean room
 and Eqn. \ref{eqn:DinfSS2} to represent the clean air
 supplied to the laboratory space outside the portable
 clean room.

Any clean room environment is likely to provide sufficient
 air flow to make aerosol infection very unlikely in the
 well-mixed approximation used in section \ref{sec:dose}.
This will leave close encounters, as described in section
 \ref{sec:close}, as the dominant infection path.
If there are multiple occupants in a clean room they should
 avoid standing close to each other, or being ``down wind''
 of each other \cite{Doyle2020}.

%%%%%%%%%%%%%%%%%%%%%%%
\subsection{UV-C Lighting}
\label{sec:UV}

Illumination of the air-space in patient rooms
 with ultra-violet light has been shown to dramatically
 reduce viral longevity in aerosol form, and thereby
 prevent infection \cite{Tellier:2006ww,UVGI_Memarzadeh:2010tk,UVGI2010}.
This could be used to decrease $\tau_a$ in spaces where
 increasing air circulation (i.e., reducing $\tauroom$)
 is impractical.
Clearly, this requires that air in the space pass
 through the UV-C illuminated volume, and that
 it received a long enough exposure to make inactivation effective.
Care is also required to avoid exposing occupants to UV-C
 radiation, which is required for viral deactivation
 but can be harmful to the eyes and skin.
This could be done geometrically (e.g., only illuminate
 spaces above \SI{2.5}{m}), or actively with motion sensors.

%%%%%%%%%%%%%%%%%%%%%%%
\section{Model Parameters}
\label{sec:parameters}

The parameters used in this model are presented in Table \ref{tab:parameters}.
All of these parameters vary between individuals and events,
 and as such the values given here are intended to serve as a means
 of making rough estimates which can guide decision making.
This section describes the provenance of these values and their
 variability.

%%%%%%%%%%%%%%%%%%%%%%%
\subsection{Viral Shedding}
\label{sec:shedding}

The volume of saliva produced in a variety of activities is used
 to understand the emission of virions into the environment
 (known as ``viral shedding''). 
The values given here are for ``typical'' individuals and behaviors,
 and actual values for any given person or event may vary by an order
 of magnitude \cite{Mittal:2020jz, duguid_1946, WHO_droplets, doi:10.1098/rsif.2013.0560, 10.1093/oxfordjournals.aje.a118097, Somsen:2020fa}.
Droplets produced while speaking, for instance, depend on speech loudness;
 speaking loudly, yelling or singing can produce an order
 of magnitude more saliva than speaking normally \cite{Stadnytskyi:2020vn}.
We are careful to avoid quantifying saliva production in terms of the
 number of droplets produced, since the large droplet-nuclei
 are relatively small in number but carry most of the virions
 (i.e., slope of the number distribution is too shallow to compensate
 the fact that volume goes with diameter cubed) \cite{duguid_1946, Morawska:2009tl}.
For a viral load of \SI{1000}{/ n L}, for example,
 a \SI{1}{\mu m} droplet nucleus has only a 1\%
 probability of containing a single virion,
 while a \SI{10}{\mu m} droplet nucleus
 is likely to contain 5--15 virions \cite{Stadnytskyi:2020vn}.

The viral load in saliva, $\rho$,
  has been seen to vary by more than 8 orders of magnitude
  in individuals that test positive for \COV\ \cite{Wolfel:2020ti, Jones2020}.
Roughly 90\% of individuals tested have a viral load less
   than $\rho_0 = \SI{1000}{/ n L}$,
   while 1\% may have a viral load above \SI{3e4}{/ n L}
   \cite{Jones2020}.
Viral load is seen to decline after symptom onset,
 so the pre-symptomatic viral load may be on the high-end
 of the distribution \cite{Wolfel:2020ti, To:2020vv, He:2020ty}.
We use $\rho_0 = \SI{1000}{/ n L}$ for our nominal dose calculations
 because it results in a rough estimate of the probability
 of infection relative to the full distribution for probabilities
 of a few percent (i.e., the ``random $\rho \Vsrc \Ninf$'' curve is
  close to the other curves around $\Dinf \ssim 0.03$
  in Fig. \ref{fig:Dinf}, right).
This choice does not impact the final probability of infection shown
 in Fig. \ref{fig:Dinf}, since a different choice of $\rho_0$ would
 shift the ``random $\rho \Vsrc \Ninf$'' curve to compensate.

%%%%%%%%%%%%%%%%%%%%%%%
\subsection{Aerosol Decay Time}
\label{sec:taua}

Ventilation rates aside, there are two timescales relevant
 to the decay of the
 infectious aerosol concentration: the decline in potency
 or ``infectivity`` of virions over time,
 and the slow settling of droplet nuclei due to gravity.
The infectivity decay time of \SCoV\ in aerosol form has been
 found to be $\taudecay \gtrsim \SI{90}{min}$
 \cite{VanDoremalen2020, Fears2020}.
Note that we are using the $1/e$ decay time, not the half-life,
 for ease of computation.

The second timescale is set by the settling
 time of the larger aerosol particles which contain the majority
 of the virions shed into the environment.
Using the ``continuous fallout'' model presented
 in \cite{Bourouiba2014} and the data presented
 in \cite{Somsen:2020fa, Stadnytskyi:2020vn} we estimate this as
\beq{taufall}
\taufall \simeq \usettle \Vroom / \Aroom \sim \SI{20}{min}
\eeq
 for a room with a \SI{3}{m} ceiling
 (i.e., the ratio of the room volume to the floor area $\Aroom$
  is \SI{3}{m}).
We use
 $\usettle \ssim \SI{0.1}{m / min} \ssim \SI{1}{mm / s}$
 for the characteristic settling speed of
 speech-generated droplet nuclei
 \cite{Stadnytskyi:2020vn, Somsen:2020fa, Johnson:2011ke},
 corresponding to a particle with an aerodynamic
 diameter of approximately $\SI{5}{\mu m}$.

Since the decay of the infectious aerosol concentration is dominated by
 settling rather the infectivity decay of the virions
 themselves, we use $\tau_a \ssim \SI{20}{min}$.
The impact of settling is non-negligible in poorly
 ventilated spaces, as demonstrated by the outbreak
 described in appendix \ref{sec:retarant}.

%%%%%%%%%%%%%%%%%%%%%%%
\subsection{Viral Intake and Infection}
\label{sec:intake}

The rate of air exchange with the environment while breathing
 (``breathing rate'' $r_b$) varies among individuals and activities
 by a factor of a few, except in the case of strenuous exercise which can increase it by as much as a factor of 5 \cite{Buonanno:2020wj}.
As such, this is a relatively well determined parameter
 and we use only the nominal value in our calculations.
Our nominal value, $r_b = \SI{10}{L / min}$,
 corresponds to a standing individual who is resting
 or engaged in light exercise,
 such that their tidal volume is about \SI{0.5}{L},
 and their breathing cycle has a \SI{3}{s} period \cite{Tidal2020}.

The respiratory infectivity, $\Ninf$, is not well known for \SCoV,
 but it has been measured for SARS \cite{Watanabe:2010el},
 other coronaviruses, and influenza \cite{TEUNIS2010215, Nikitin:2014ua}.
Variation by an order of magnitude among individuals is observed
 both for coronaviruses and influenza.
The number we use, $\Ninf = 1000$,
 is intended to be ``typical'' for coronaviruses
 \cite{Barr:2020ui}
 and represents roughly 100 ``plaque forming units'' (PFU)
 as measure for SARS \cite{Watanabe:2010el},
 each of which is roughly 10 virions
 \cite{Fonville:2015in}.

%%%%%%%%%%%%%%%%%%%%%%%
\section{Transmission Probability}
\label{sec:ptrn}

In this section we relate the infectious dose calculations
 presented in section \ref{sec:dose} to the probability
 of infection (for a single individual) or transmission
 (to any member of a group) while accounting for
 the probability distributions of the parameters
 discussed in section \ref{sec:parameters}
 and appendix \ref{sec:pdfs}.

While the infectious dose for any set of parameters
 can be estimated using the equations in section \ref{sec:dose},
 estimating the probability of infection in a given
 situation requires marginalizing over the probability
 distribution function associated with each of the parameters.
Fortunately, marginalization is made simpler by
 the fact that $\Dinf$ is proportional to $\rho \Vsrc / \Ninf$
 in all dose calculations, allowing us to connect
 $\Dinf$ computed from parameters in Table \ref{tab:parameters}
 to $\Pinf$ by introducing the concept of a
 ``nominal infectious dose'' $\DinfNom$.
Figure \ref{fig:Dinf} (left) shows the
 cumulative distribution of $\Dinf$
 relative to the nominal value $\DinfNom$ obtained using the
 parameters in Table \ref{tab:parameters}.

Similarly, the ``random $\rho \Vsrc \Ninf$''
 curve in Fig. \ref{fig:Dinf} (right) can be used to
 estimate the probability of a contagious individual causing
 infection in the people they interact with,
 given a nominal infectious dose $\DinfNom$
 computed with the parameters in Table \ref{tab:parameters}.
For instance, using Fig. \ref{fig:Dinf} (right, blue curve)
 we can estimate that a brief encounter which
 delivers $\DinfNom \ssim 0.01$
 results in $\Pinf \ssim 2\%$.
This infection probability accounts for parameter
 variation relative to the nominal values,
 so no further computation is required.
 
The infectious dose can be summed over multiple encounters
 to compute the probability of infection in at least one
 susceptible individual.
For example, if a contagious individual delivers a
 dose of $\DinfNom \ssim 0.01$
 on each of 10 encounters with susceptible individuals,
 then the total infectious dose is $\DinfNom \sim 0.1$
 and the probability of transmission to at least one person
 is $\Pinf \ssim 6\%$
 (again using Fig. \ref{fig:Dinf}, right).
The distribution of virions via a building's HVAC
 system, discussed in section \ref{sec:HVAC},
 provides a similar example of a
 weak interaction with a large group of people (occupants of rooms
 connected via the HVAC system).
 
Note that the final probability of transmission is \emph{not} the
 sum of the infection probabilities for each susceptible individual
 because of the correlation between these probabilities
 (i.e., the viral load of the contagious individual is common
 to all events).
It is exactly this correlation that leads to large outbreaks:
 if a highly-contagious individual delivers
 an infectious dose $\Dinf \ssim 2$ to many individuals they
 will each have a 80\% probability of being infected
 (``random $\Ninf$'' curve in Fig. \ref{fig:Dinf}, right). 

The above example shows that limiting the number of
 potential transmission events between members of a group,
 and limiting group size, are both important to minimizing
 transmission and preventing outbreaks.
If only 4 encounters are allowed per person,
 and the contagious individual only encounters 2 people,
 then $\DinfNom \lesssim 0.05$ per encounter gives
 a total $\DinfNom \lesssim 0.4$ and
 a 10\% probability of transmission.

%%%%%%%%%%%%%%%%%%%%%%%
\section{Conclusions}
\label{sec:conclusions}

The new world of \COV\ is here and we will all have to learn
 to live in it.
Understanding how to navigate the dangers of this new world
 will be necessary as we come out of our
 houses and return to our workplaces.
The calculations presented here suggest that keeping
 the risk of infection low in the workplace may require
 mitigation techniques and protocols that limit the infectious
 dose a contagious individual can deliver to
 their coworkers.

Some broad guidelines which we draw from the above analysis are:
\begin{enumerate}[leftmargin=*]
    \item Aerosol build-up in closed spaces should be treated with care. Avoiding infection requires good ventilation and/or short exposure times.  Generally, office spaces should not be occupied by
     more than one person.
     In the early phases of epidemic decay,
     airflow in shared spaces should be at least
     $\SI{50}{m^3 / min} \ssim \SI{2000}{CFM}$
     per occupant beyond the first.
     As the prevalence of \COV\ in the population decreases this
     could be reduced as low as
     $\SI{5}{m^3 / min} \ssim \SI{200}{CFM}$
     per occupant beyond the first (see section \ref{sec:steady}).
    \item Recirculation in HVAC systems should be avoided if possible.
     High quality filters (e.g., MERV 12) should be used
     in recirculating HVAC systems,
     and office use should be kept to a minimum
     to avoid transmission through the air circulation.
     In the early phases of epidemic decay, airflow should be at least
     $\SI{10}{m^3 / min} \ssim \SI{350}{CFM}$
     per occupant (averaged over the full HVAC system distribution,
     see section \ref{sec:HVAC}).
    \item Small or poorly ventilated common spaces
     where many people spend time
     (i.e., bathrooms, elevators and stairwells) are of particular concern.
     Consider increasing airflow and/or adding local air scrubbers to
      avoid wait times between occupants.
     For single occupancy shared spaces,
     $\SI{10}{m^3 / min} \ssim \SI{350}{CFM}$
     should be considered a minimum (see section \ref{sec:transient}).
    \item Mask use may help to prevent direct exposure when
     a minimum of \SI{2}{m} interpersonal distance
     cannot be maintained, but is not sufficient to prevent infection
     in an enclosed space
     \emph{regardless of the distance between occupants}.
     Distances less than \SI{1}{m} remain more dangerous than larger
     distances due to aerosol leakage around masks,
     especially in the event of coughing or sneezing.
     Additional personal protective equipment
     (PPE) should be used for tasks which require close proximity
     (see sections \ref{sec:close} and \ref{sec:masks}).
\end{enumerate}

%%%%%%%%%%.%%%%%%%%%%%%%
\section{Acknowledgements}
\label{sec:ack}

The author would like to thank Lisa Barsotti,
 Paola Rebusco, Rich Abbott,
 and Paola Cappellaro for their careful reading
 and editing of this article.
The analysis presented here would not have
 been possible without the inspiration
 provided by Dawn Mautner,
 the information supplied by Scott Kemp
 and Peter Fisher,
 and the many hours of quiet time for which
 I have Lisetta Turini to thank.
An early draft of \cite{Doyle2020} and feedback from its
 authors were both very helpful and much appreciated.
Subsequent feedback from Lisa Brosseau 
 and Dave Boggs has
 been critical to deepening our literature base
 and refining our presentation.
The author is, as always, very grateful for the computing
 support provided by The MathWorks, Inc.

%%%%%%%%%%%%%%%%%%%%%%%
\end{multicols}

%\vspace{5ex}
\newpage
\appendix
%%%%%%%%%%%%%%%%%%%%%%%

%%%%%%%%%%%%%%%%%%%%%%%
\section{Outbreaks}
\label{sec:outbreaks}
\vskip -0.05in

Documented outbreaks offer a means for checking the plausibility of
 our aerosol transmission model.
In this section we select a few outbreaks where transmission
 via aerosols appears likely (e.g., distances were too large
 for droplets, or the pattern of infection followed the airflow).
%While surface contamination cannot be ruled out, each of these cases 

Information about the outbreaks is, however, limited,
 and the viral load of the contagious individual
 is not known, so these comparisons only provide weak constraints
 on the model parameters.
In particular, reported outbreaks are likely to involve
 unusually contagious individuals (a.k.a. ``super-spreaders''),
 with $\rho_0 \gtrsim \SI{1000}{/ n L}$
 (roughly 10\% of cases), so we use that value
 in these computations.
Since this fixes the value of $\rho$, we use the
 ``random $\Vsrc \Ninf$'' curve in Fig. \ref{fig:Dinf} to compute
 the probability of infection.

%%%%%%%%%%%%%%%%%%%%%%%
\subsection{Guangzhou Restaurant}
\label{sec:retarant}
\vskip -0.05in

\cite{Li:2020um, Lu:2020tw} document a COVID-19 outbreak associated
 with air flow in a restaurant (Guangzhou, China).
Unlike many outbreaks, this one was studied in great
 detail, including on-site experimental recreation of the
 airflow and computational fluid dynamics (CFD) modeling,
 making it an excellent cross-check for our
 simplified aerosol model \cite{Li:2020um}.
 
In this outbreak, there was one contagious individual (CI)
 seated at a table with 9 family members.
Air circulation in the restaurant was dominated by wall
 mounted air conditioning units, one of which essentially
 defined a well mixed air space that included 2
 other families totaling 11 people.
The distance between CI and the people infected
 varied from $\ssim \SI{50}{cm}$ to $\SI{4.6}{m}$
 with 4 of the 5 people who were infected at other
 tables 2 or more meters from CI \cite{Li:2020um}.
There is no obvious correlation between who was infected
 and their distance from CI or the direction that CI
 was facing.
 
Of the people at the table with CI, 4 of 9 were infected,
 while 5 of the 11 people at the other 2 tables were infected
 (45\% attack rate in both cases).
The aerosol concentration at these three tables was found
 to be quite similar, both experimentally
 and in CFD models \cite{Li:2020um}.
Given the common source (i.e., the contagious individual),
 we use the ``random $\Ninf$'' curve in Fig. \ref{fig:Dinf}
 (right, yellow) to estimate the infectious dose required
 to produce this attack rate as $\Dinf \simeq 0.6$.
In the next paragraph we compare this value with the estimate
 we get from direct calculation using Eqn. \ref{eqn:DinfSS}. 

The recirculating air space defined by the wall mounted AC
 unit which transported air from CI
 to other diners who were infected is roughly \SI{60}{m^3}
 and was found to have $\rroom \ssim \SI{1}{m^3/min}$,
 and roughly 1 air change per hour
 ($\tauroom \ssim \SI{60}{min}$).
The overlap in time between CI and the other diners
 in this air space was approximately 60 minutes.
This is situation is not in the well ventilated
 limit, and has $f_a \ssim 1/4$ according to Eqn. \ref{eqn:fa}
 with our estimated 20 minute time-constant for settling
 (see section \ref{sec:taua}).

Assuming a somewhat talkative contagious individual
 ($\rsrc \ssim \SI{4}{n L / min}$),
 and given $\rroom \ssim \SI{1}{m^3/min}$ and
 $t \ssim \SI{60}{min}$,
 the other diners were exposed to 
 a \emph{nominal} infectious dose of $\DinfNom \ssim 0.6$.
This is a perfect match to
 the value computed using the attack rate,
 and indicates that CI's viral load was likely
 around $\rho_0$ at the time of this outbreak.
This is not surprising since, as noted above,
 there is a selection bias when
 working with noteworthy outbreaks
 that precludes low values of $\rho$.

The detailed analysis of this outbreak offers a number of
 other quantitative and qualitative tests of our
 simplified model of aerosol infection.
\cite{Li:2020um} notes that a table near the CI
 (``table 17''), but not in the air space defined by the AC unit,
 had a relatively high aerosol concentration due to leakage
 from the primary air circulation pattern.
There were 5 people at this table, and they
 overlapped with CI for 18 minutes at the end of CI's stay.
It is unlikely that the infectious dose
 summed over all 5 diners was larger than 1,
 since $\Pinf \simeq 0.7$ for $\Dinf = 1$,
 indicating an average $\Dinf \lesssim 0.2$
 with 70\% confidence (and average $\Dinf \lesssim 0.6$
 with 90\% confidence).
This is consistent with the shorter overlap time
 and the somewhat reduced aerosol concentration at
 table 17.
 
However, a similar analysis applied to the 50 diners
 who were at more distant tables
 (e.g., ``table 10'' in \cite{Li:2020um}), none of whom
 were infected, indicates a relatively short aerosol decay time.
Since the total infectious dose required to cause at least
 one infection does not depend on the number of people
 exposed
 (i.e., total $\Dinf \lesssim 1$ with 70\% confidence,
 and $\Dinf \lesssim 3$ with 90\% confidence),
 the average infectious dose for these 50 diners
 must have been less than $\Dinf \lesssim 0.06$
 (90\% confidence).
Assuming an average overlap time with CI of at least
 30 minutes, the virion aerosol concentration must
 have been \emph{at least 5 times less} than the concentration
 around the 3 tables where the attack rate was high.
Using the data in figure S4 of \cite{Li:2020um},
 we can see that this requires the aerosol concentration
 to come to steady-state much more quickly than the
 tracer gas used in that study.
Steady state was likely achieved in less than 10 minutes,
 and certainly in less than 20 minutes.
This agrees with the settling times of \SIrange{5}{10}{\mu m}
 droplet nuclei, as discussed in section \ref{sec:taua}.

Analysis of surveillance videos reveals no close contact,
 or objects shared, among members of different families.
None of the restaurant staff were infected, nor were
 patrons of other businesses in the building.
Simply put, this outbreak is essentially impossible to explain with
 surface contamination or large droplet driven transmission,
 and yet easily understood in terms of aerosol transmission.

%%%%%%%%%%%%%%%%%%%%%%%
\subsection{Hunan Coach}

\cite{BusOutbreak} documents an outbreak on a
 long distance coach (Hunan, China).
This 45 person coach should have $\rroom \simeq \SI{8}{m^3 / min}$
 \cite{Qian2020.04.04.20053058},
 and we will estimate the ride duration as 2 hours.
The contagious individual did not interact with others,
 so we assign a source rate of $\rsrc = r_b = \SI{1}{n L / min}$.
The resulting nominal dose of $\DinfNom \ssim 0.15$ is in reasonable
 agreement with the fact that 8 of the 45 passengers were infected.
The infections were somewhat localized,
 with the most distant 4.5 meters from the contagious individual,
 possibly due to incomplete mixing of the air in the bus.
 
This outbreak also resulted in the infection of a passenger who
 boarded 30 minutes after the contagious passenger disembarked.
This could have been due to surface contamination,
 but then one must wonder why no passengers on later voyages
 were infected (since fomites last for days) \cite{VanDoremalen2020}.
More likely it indicates settling and air exchange
 while the bus was stopped
 resulted in an order of magnitude reduction in infectious dose.
(Outbreaks on vehicles were common in China,
 possibly due to poor ventilation \cite{Qian2020.04.04.20053058}.)

%%%%%%%%%%%%%%%%%%%%%%%
\subsection{Seattle Choir}

\cite{Hamner:2020xx, ChoirOutbreak}
 Of 60 singers 87\% infected after singing together for 2.5 hours in a small church ($\Vroom \ssim \SI{600}{m^3}$).
An air-cycle time of 30 minutes gives
 $\rroom \ssim \SI{20}{m^3/min}$, and singing
 can be approximated by $\rsrc \ssim \SI{30}{n L / min}$.
The 150 minute exposure yields an infectious dose of
 $\Dinf \ssim 0.9$ (accounting for settling with $f_a = 0.4$).
Sadly, this makes the probability of infection quite high
 ($\Pinf \ssim 60\%$).
Even more sadly, this is not the only choir outbreak
 \cite{ChoirOutbreak2, ChoirOutbreak3},
 and this can happen even if distancing and hand-hygiene
 rules are strictly followed \cite{ChoirOutbreak4}.
The least we can do is to learn from their tragic experience.
(Like choir practice, aerobic dance classes \cite{Jang:2020tu}
 and call centers \cite{Park:2020tv} are ideal environments
 for \SCoV\ transmission.)

%%%%%%%%%%%%%%%%%%%%%%%
\subsection{Diamond Princess}

\cite{Xu2020.04.09.20059113} concludes from the outbreak on
 the Diamond Princess cruise ship that
 long-range airborne transmission is unlikely since \SCoV\ did
 not pass through the ship's air conditioning system.
Two Okinawa taxi-drivers were, however, infected by passengers
 during a shore visit.
This outbreak is different from the others in that the
 contagious individual who initiated the outbreak
 was not involved in the taxi-driver infections.
At the time of the shore visit in Okinawa there were
 roughly 25 pre-symptomatic infected passengers on the
 Diamond Princess at least a few of whom disembarked.
 
If we assume that the taxi ride lasted 10 minutes
 (the port in Naha is close to the main attractions),
 that the contagious passenger spent the ride talking
 to another passenger (or coughed once),
 and that the taxi had the fan on low ($\rroom \ssim \SI{1}{m^3/min}$),
 the driver was exposed to an infectious dose of
 $\Dinf \ssim 1$ according to Eqn. \ref{eqn:DinfSS}.
Using the ``random $\rho \Vsrc \Ninf$'' curve in Fig. \ref{fig:Dinf}
 since these were secondary infections, we find that
 the probability of transmission was about 15\%
 for each contagious passenger who disembarked.
This probability is relatively insensitive to our assumptions
 and would only change by a factor of 2 if the dose changed
 by an order of magnitude (in either direction).
If both infections in Naha were caused by the same
 contagious individual
 (one on the ride in, one on the ride out)
 it would be appropriate to use the ``random $\Vsrc \Ninf$''
 probability of 60\%.

Fortunately, there were no other \COV\ infections
 linked to the Diamond Princess' stop in Okinawa,
 indicating that the closed environment and long exposure time
 of the taxi was likely a key ingredient for transmission.
That is, changing the air-flow rate in the above calculation
 to $\rroom \gtrsim \SI{100}{m^3/min}$ to represent
 shops and more open spaces would reduce the probability
 of infection to less than 2\%.
And, while surface transmission could explain transmission
 to taxi drivers, it does not explain the absence of transmission
 to shop keepers or other people with whom the passengers
 of the Diamond Princess interacted.

%%%%%%%%%%%%%%%%%%%%%%%
\subsection{Hospital Air Sampling}

There was no outbreak in the Nebraska Medical Center,
 but air sampling in and around \COV\ patient rooms
 \cite{Santarpia2020.03.23.20039446}
 offers a further cross-check of the calculations
 presented here.
Approximately 3 viral RNA copies per liter
 of air sampled were found in a patient's room (and in the
 hallway outside the patient's room after the door was opened). 
These rooms have $\tauroom \simeq \SI{8}{min}$ and
 $\Vroom \simeq \SI{30}{m^3}$, indicating an airflow
 rate of $\rroom \ssim \SI{4}{m^3 / min}$ \cite{Hewlett:2013tt}.
Equation \ref{eqn:rhoSS} indicates a steady-state concentration
 of $\rhoASS \simeq \SI{0.25}{/ L}$.
This suggests that either
 this patient had a very high viral load ($\rho \sim \SI{1e10}{/ mL}$,
 which is in the top 3\%), or that they were occasionally
 talking or coughing and had $\rho \ssim \rho_0$.

Air sampling data is also available from hospitals and public areas in China.
\cite{Liu2020.03.08.982637} reports quantitative viral
 deposition rate of roughly \SI{2}{/ m^2} per minute
 in a patient's room.
The area sampled was \SI{3}{m} from the patient's bed,
 so too far for most droplets \cite{Bourouiba2014}.
If we assume that this is due to slow settling of the larger
 droplet-nuclei with characteristic a speed of
 $\usettle \ssim \SI{0.1}{m / min}$
 (see section \ref{sec:taua}),
 this implies a aerosol concentration of roughly
 \SI{20}{/ m^3}.
(Oddly, air samplers in patient rooms did not detect concentrations
 above their detection threshold, but the patient's bathroom and
 other areas in the hospital had concentrations near \SI{20}{/ m^3}).

A concentration of \SI{20}{/ m^3}
 is more than a factor of 100 lower than that reported in 
 Nebraska \cite{Santarpia2020.03.23.20039446}, but still implies
 $\rho$ above the median of the distribution
 function shown in Fig. \ref{fig:PDFs}.
If we assume an air flow rate in the room of
 $\rroom \ssim \SI{4}{m^3 / min}$ and
 viral shedding dominated by breathing (i.e., $\Vsrc \ssim V_b$),
 for instance, $\rho$ is roughly $\SI{1e8}{/ m L} = \rho_0 / 10$.

%%%%%%%%%%%%%%%%%%%%%%%
\section{Parameter Probability Distributions}
\label{sec:pdfs}

As described in section \ref{sec:parameters} several parameters
 used in this work vary significantly between individuals and
 events.
This appendix describe the details of the probability
 density functions use for $\rho$, $\Vsrc$ and $\Ninf$.

\begin{figure*}[h!]
    \label{fig:PDFs}
 \begin{center}
 \includegraphics[width=0.48\textwidth, trim=15mm 5mm 15mm 15mm, clip]{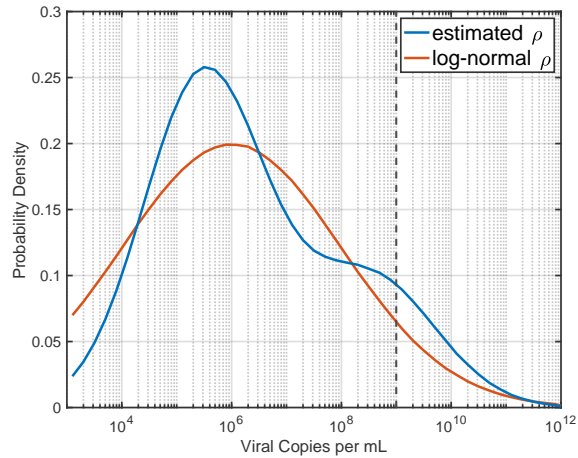}
 \end{center}
 \caption{Probability density functions for $\rho$.
 The estimated distribution used in this work is shown along with
  a log-normal distribution centered at \SI{1e6}{/ m L}.
 The vertical line at \SI{1e9}{/ m L} indicates our
  ``nominal $\rho_0$'' of \SI{1000}{/ n L}.
  10\% of the estimated $\rho$ distribution lies above this line
   and 7\% of the log-normal distribution lies above it.}
\end{figure*}

Figure \ref{fig:PDFs} shows the distribution for $\rho$.
The estimated distribution used in this work is derived
 from \cite{Jones2020} with the low end of the distribution
 pushed up somewhat to account for the downward trend in
 viral load after symptom onset.
The log-normal distribution centered at \SI{1e6}{/ mL}
 with $\sigma = 10^2$ is estimated from the linear fit
 in figure 2 of \cite{To:2020vv}.
These distributions both have a median viral load of
 \SI{1e6}{/ m L} and roughly 10\% of cases above \SI{1e9}{/ m L}.
We have computed $\Pinf$ as in Fig. \ref{fig:Dinf} for
 both distributions and they give quantitatively similar, 
 and qualitatively identical, results.

To account for individual variability, we use a log-normal
 distribution for $\Ninf$ which is centered around 1000
 and has a width of a factor of 2.
This makes the 95\% confidence interval 250--4000
 which is in reasonable agreement with \cite{Watanabe:2010el}
 and \cite{TEUNIS2010215}.
 
Similarly, an order of magnitude variation among
 individuals and events is also expected
 in droplet and aerosol production
 \cite{doi:10.1098/rsif.2013.0560, Stadnytskyi:2020vn}.
For this we also use a log-normal distribution centered
 around the $\Vsrc$ values given in Table \ref{tab:parameters}
 with a width of a factor of 3.

%%%%%%%%%%%%%%%%%%%%%%%
% [trim=left bottom right top, clip]

%%%%%%%%%%%%%%%%%%%%%%%
%%%%%%%%%%%%%%%%%%%%%%%
%%%%%%%%%%%%%%%%%%%%%%%
\vspace{10ex}

\bibliographystyle{unsrt}  
\bibliography{references} 

\end{document}